\documentclass[article,preprint,groupedaddress]{revtex4}
\usepackage{epsfig}
\usepackage{graphicx}

\begin{document}

\title{Lower bound on the radii of black-hole photonspheres}
\author{Shahar Hod}
\address{The Ruppin Academic Center, Emeq Hefer 40250, Israel}
\address{}
\address{The Hadassah Institute, Jerusalem 91010, Israel}
\date{\today}

\begin{abstract}

\ \ \ The existence of closed null circular geodesics around black
holes is one of the most intriguing predictions of general
relativity. It has recently been conjectured that the radii of
black-hole photonspheres are bounded from below by the simple
relation $r_{\text{ph}}\geq {3\over2}r_{\text{H}}$, where
$r_{\text{H}}$ is the radius of the outer black-hole horizon. We
here prove the validity of this conjecture for spherically symmetric
hairy black-hole configurations whose radial pressure function
$P\equiv |r^3p|$ decreases monotonically.
\end{abstract}
\bigskip
\maketitle

\section{Introduction}

Null circular geodesics, closed orbits on which massless particles
(photons, gravitons) can orbit a central black hole, provide
valuable information about the physical characteristics of the
corresponding curved spacetime \cite{Bar,Chan,Shap,Hodns}. Due to
their importance in astrophysical \cite{Pod,Ame}, cosmological
\cite{Ste}, and theoretical
\cite{Mash,Goeb,Hod1,Dec,Hodhair,Hodfast,YP,Hodub,Lu1}
studies of black-hole
spacetimes , light-like circular geodesics have attracted over the years a good deal of
attention from physicists and mathematicians.

In an astrophysical context, the photonsphere (a
compact null hypersurface around the central black hole) determines the optical appearance of a
compact collapsing object as seen by external asymptotic observers \cite{Pod,Ame}.
Likewise, the intriguing general relativistic phenomenon of strong
gravitational lensing by black holes is closely related to the presence of
null circular geodesics in these curved spacetimes \cite{Ste}.

In addition, the physical properties of unstable null circular geodesics in black-hole
spacetimes are known to determine the complex resonant spectra (the
quasinormal frequencies) that characterize the corresponding curved spacetimes in
the eikonal (geometric-optics) regime (see \cite{Mash,Goeb,Hod1,Dec}
and references therein).

Interestingly, it has been proved in \cite{Hodhair} that, in
spherically symmetric hairy black-hole spacetimes, the radius
$r_{\gamma}$ of the innermost null circular geodesic sets a lower
bound on the effective length of the external non-linear matter
fields. In particular, it has been proved that in curved black-hole spacetimes that possess hair,
the effective radius of the hair is bounded from below by the compact relation
$r_{\text{hair}}\geq r_{\gamma}$ \cite{Hodhair}.

In addition, it has been proved that, in spherically symmetric
\cite{Hodfast} as well as in axisymmetric \cite{YP}
black-hole spacetimes, the innermost null circular
orbit provides the fastest way to circle the central compact black
hole as measured by asymptotic observers.

Using analytical techniques, it has been proved in \cite{Hodub}
that, for spherically symmetric hairy black-hole spacetimes whose
external matter fields are characterized by a non-positive
energy-momentum trace \cite{Bond1}, the radius $r_{\gamma}$ of the innermost null circular
geodesic of the curved black-hole spacetime is bounded from {\it
above} by the remarkably compact relation \cite{Noteunit}
\begin{equation}\label{Eq1}
r_{\gamma}\leq 3M\  ,
\end{equation}
where $M$ is the total ADM mass of the spacetime.

Intriguingly, based on the analysis of the physical properties of
some non-trivial black-hole spacetimes, it has recently been
conjectured \cite{Lu1}
that the radii of black-hole photonspheres
are bounded from {\it below} by the simple relation
\begin{equation}\label{Eq2}
r_{\gamma}\geq {3\over2}r_{\text{H}}\  ,
\end{equation}
where $r_{\text{H}}$ is the radius of the outer black-hole horizon.

The main goal of the present compact paper is to explore the regime of
validity of this physically interesting conjecture. In particular,
below we shall explicitly prove that the null circular geodesics of
spherically symmetric hairy black-hole spacetimes whose radial
pressure function $P\equiv |r^3p|$ decreases monotonically are
characterized by the conjectured \cite{Lu1} lower bound (\ref{Eq2}).

\section{Description of the system}

We shall analyze the null circular geodesics of
spherically-symmetric non-vacuum black-hole spacetimes. The curved
line element can be expressed in the form
\cite{Hodfast,Hodm,Notesch}
\begin{equation}\label{Eq3}
ds^2=-e^{-2\delta}\mu dt^2 +\mu^{-1}dr^2+r^2(d\theta^2 +\sin^2\theta
d\phi^2)\  ,
\end{equation}
where $\delta=\delta(r)$ and $\mu=\mu(r)$ are the radially-dependent metric functions.

In terms of the line element (\ref{Eq3}), the Einstein differential
equations $G^{\mu}_{\nu}=8\pi T^{\mu}_{\nu}$ can be expressed in the form \cite{Hodfast,Hodm,Noteprim}
\begin{equation}\label{Eq4}
\mu'=-8\pi r\rho+(1-\mu)/r\
\end{equation}
and
\begin{equation}\label{Eq5}
\delta'=-4\pi r(\rho +p)/\mu\  .
\end{equation}
Here $\rho\equiv-T^{t}_{t}$, $p\equiv T^{r}_{r}$, and
$p_T\equiv T^{\theta}_{\theta}=T^{\phi}_{\phi}$, where $\rho$, $p$, and
$p_T$ are respectively the energy density, radial pressure, and
tangential pressure of the external matter fields \cite{Bond1}.

Regularity of the spacetime at the black-hole outer horizon,
$r=r_{\text{H}}$, enforces the boundary conditions \cite{Bekreg}
\begin{equation}\label{Eq6}
\mu(r_{\text{H}})=0\ \ \ {\text{with}}\ \ \ \mu'(r_{\text{H}})\geq
0\
\end{equation}
and
\begin{equation}\label{Eq7}
\delta(r_{\text{H}})<\infty\ \ \ ; \ \ \
\delta'(r_{\text{H}})<\infty\ .
\end{equation}
From Eqs. (\ref{Eq5}), (\ref{Eq6}), and (\ref{Eq7}) one finds the simple boundary condition \cite{Bekreg}
\begin{equation}\label{Eq8}
p_{\text{H}}=-\rho_{\text{H}}\
\end{equation}
at the black-hole horizon, where $p_{\text{H}}\equiv
p(r=r_{\text{H}})$ and $\rho_{\text{H}}\equiv \rho(r=r_{\text{H}})$. In
addition, from Eqs. (\ref{Eq4}), (\ref{Eq6}), and (\ref{Eq8}) one obtains the
boundary condition \cite{Bekreg,Noteext}
\begin{equation}\label{Eq9}
-8\pi r^2_{\text{H}}p_{\text{H}}\leq 1\
\end{equation}
for the radial pressure function.

We shall assume that the components of the energy-momentum tensor
satisfy the well-known dominant energy condition, according to which
the energy density of the matter fields is positive semidefinite and
it bounds the pressure components \cite{Bekreg}:
\begin{equation}\label{Eq10}
\rho\geq 0\ \ \ \ ; \ \ \ \ \rho\geq |p|,|p_T|\  .
\end{equation}
In addition, following \cite{Bond1} we shall assume that the external matter
fields of the black-hole spacetime are characterized by a
non-positive trace of the energy-momentum tensor:
\begin{equation}\label{Eq11}
T\leq 0\  ,
\end{equation}
where $T=-\rho+p+2p_T$.

For later purposes we note that the gravitational mass $m(r)$
contained within a sphere of radius $r$ is given by the integral
relation
\begin{equation}\label{Eq12}
m(r)={1\over2}r_{\text{H}}+\int_{r_{\text{H}}}^{r} 4\pi r'^{2}
\rho(r')dr'\  .
\end{equation}
Here $m(r_{\text{H}})=r_{\text{H}}/2$ is the mass contained within
the black-hole horizon. Taking cognizance of Eqs. (\ref{Eq4}) and (\ref{Eq12}),
one can express the radially-dependent metric function $\mu(r)$ in terms of the
mass function $m(r)$:
\begin{equation}\label{Eq13}
\mu(r)=1-{{2m(r)}\over{r}}\  .
\end{equation}

\section{The lower bound on the radii of black-hole photon-spheres}

In the present section we shall consider the following physically
interesting question: How close can the innermost null circular
geodesic of a black-hole spacetime be to its outer horizon? Below we
shall explicitly prove that in hairy black-hole spacetimes whose
radial pressure function \cite{Notenp}
\begin{equation}\label{Eq14}
P\equiv |r^3p(r)|\
\end{equation}
decreases monotonically \cite{Notemov,Notehg}, the null circular
geodesics cannot lie arbitrarily close to the outer black-hole
horizon. In particular, we shall show that the null circular
geodesics of these black-hole spacetimes are characterized by the
lower bound (\ref{Eq2}).

We shall first derive a lower bound on the mass $m_{\text{hair}}$ of
the matter fields (hair) outside the black-hole horizon. Taking cognizance
of Eqs. (\ref{Eq10}) and (\ref{Eq12}), one finds the inequality
\begin{equation}\label{Eq15}
m_{\text{hair}}=\int_{r_{\text{H}}}^{\infty} 4\pi r^2\rho(r)dr\geq
-\int_{r_{\text{H}}}^{\infty} 4\pi r^2p(r)dr\ .
\end{equation}

Interestingly, and most importantly for our analysis, it has been
proved in \cite{Hodhair} that the pressure function $r^4p$ of hairy matter fields
which satisfy the energy conditions (\ref{Eq10}) and (\ref{Eq11})
is non-positive and monotonically decreasing in the interval
$r\in [r_{\text{H}},r_{\gamma}]$:
\begin{equation}\label{Eq16}
\{p(r)\leq0\ \ \ \text{and}\ \ \ (r^4p)'\leq0\}\ \ \ \ \text{for}\ \
\ \ r_{\text{H}}\leq r\leq r_{\gamma}\  .
\end{equation}
In particular, from (\ref{Eq16}) one deduces the simple relation
\begin{equation}\label{Eq17}
0\leq -r^4_{\text{H}}p_{\text{H}}\leq -r^4p(r)\ \ \ \ \text{for}\ \
\ \ r_{\text{H}}\leq r\leq r_{\gamma}\  .
\end{equation}
Using the relation (\ref{Eq17}), one obtains from (\ref{Eq15}) the series of inequalities
\begin{equation}\label{Eq18}
m_{\text{hair}}\geq
-\int_{r_{\text{H}}}^{r_{\gamma}} 4\pi r^2p(r)dr
\geq-\int_{r_{\text{H}}}^{r_{\gamma}} 4\pi
{{r^4_{\text{H}}p_{\text{H}}}\over{r^2}}dr=-4\pi
r^4_{\text{H}}p_{\text{H}}\Big({{1}\over{r_{\text{H}}}}-{{1}\over{r_{\gamma}}}\Big)\
\end{equation}
for the mass $m_{\text{hair}}$ of the external matter fields (hair).

Using the Einstein field equations (\ref{Eq4}) and (\ref{Eq5}), it has
been explicitly proved \cite{Hodhair} that the black-hole null
circular geodesics are characterized by the relation \cite{Notenex}
\begin{equation}\label{Eq19}
{\cal N}(r)\equiv 3\mu(r)-1-8\pi r^2p(r)=0\ \ \ \ \text{for}\ \ \ \
r=r_{\gamma}\  .
\end{equation}
Substituting the lower bound (\ref{Eq18}) into Eq. (\ref{Eq19}) and using the relation (\ref{Eq13}), one
obtains the inequality
\begin{equation}\label{Eq20}
r_{\gamma}-{3\over2}r_{\text{H}}+12\pi
r^4_{\text{H}}p_{\text{H}}\Big({{1}\over{r_{\text{H}}}}-{{1}\over{r_{\gamma}}}\Big)-4\pi
r^3_{\gamma}p_{\gamma}\geq0\  ,
\end{equation}
which characterizes the null circular geodesics of the
spherically-symmetric hairy black-hole spacetime. In addition, using the
inequality $r^3_{\text{H}}p_{\text{H}}\leq r^3_{\gamma}p_{\gamma}$,
which follows from the assumed
monotonic behavior of the radial pressure function (\ref{Eq14}) \cite{Notehg} and the
fact that the radial pressure is non-positive between
the black-hole horizon and the innermost null circular geodesic [see Eq. (\ref{Eq16})], one deduces
from (\ref{Eq20}) the inequality
\begin{equation}\label{Eq21}
r_{\gamma}-{3\over2}r_{\text{H}}+12\pi r_{\text{H}}
r^3_{\gamma}p_{\gamma}\Big({{1}\over{r_{\text{H}}}}-{{1}\over{r_{\gamma}}}\Big)-4\pi
r^3_{\gamma}p_{\gamma}\geq0\  ,
\end{equation}
which can be expressed in the remarkably compact form
\begin{equation}\label{Eq22}
\Big(r_{\gamma}-{3\over2}r_{\text{H}}\Big)\cdot(1+8\pi
r^2_{\gamma}p_{\gamma})\geq0\  .
\end{equation}

Taking cognizance of the characteristic relation (\ref{Eq19}) for
the null circular geodesics of the black-hole spacetime
(\ref{Eq3}), one can write (\ref{Eq22}) in the form
\begin{equation}\label{Eq23}
\Big(r_{\gamma}-{3\over2}r_{\text{H}}\Big)\cdot\mu(r_{\gamma})\geq0\
,
\end{equation}
which yields the lower bound
\begin{equation}\label{Eq24}
r_{\gamma}\geq{3\over2}r_{\text{H}}\
\end{equation}
on the radii of the black-hole null circular geodesics.

\section{Summary}

Black-hole spacetimes are characterized by the presence of
null circular geodesics on which massless particles (photons, gravitons) can orbit
the central black hole. These closed light-like orbits play important physical
roles in astrophysical \cite{Pod,Ame}, cosmological \cite{Ste}, and
theoretical \cite{Mash,Goeb,Hod1,Dec,Hodhair,Hodfast,YP,Hodub,Lu1}
studies
of curved black-hole spacetimes.

In the present compact paper we have addressed the following physically
interesting question: In a spherically symmetric black-hole
spacetime, how close can the innermost null circular geodesic be to
the black-hole horizon? Interestingly, it has recently been
conjectured \cite{Lu1}
that, for a black hole of horizon radius
$r_{\text{H}}$, massless particles can orbit the central black hole
on circular trajectories whose radii are bounded from below by the
simple relation $r_{\gamma}\geq {3\over2}r_{\text{H}}$ [see Eq.
(\ref{Eq2})].

Using {\it analytical} techniques, we have presented a remarkably
compact theorem that proves the validity of this intriguing
conjecture for spherically symmetric hairy black-hole spacetimes
whose radial pressure function $|r^3p(r)|$, which characterizes the
spatial behavior of the external matter fields (hair), decreases
monotonically.

\newpage

\bigskip
\noindent {\bf ACKNOWLEDGMENTS}

This research is supported by the Carmel Science Foundation. I thank
Yael Oren, Arbel M. Ongo, Ayelet B. Lata, and Alona B. Tea for
stimulating discussions.


\begin{thebibliography}{99}
\bibitem{Noteunit} We use gravitational units in which $G=c=1$.
\bibitem{Notesch} Here $\{t,r,\theta,\phi\}$ are the Schwarzschild spacetime coordinates.
\bibitem{Notenp} It is worth emphasizing that the radial pressure
$p(r)$ is non-positive in the interval $r_{\text{H}}\leq r\leq
r_{\gamma}$ between the outer black-hole horizon and the innermost null circular geodesic
[see \cite{Hodhair} and Eq. (\ref{Eq16}) below].
\bibitem{Notemov} It is worth noting that the physical motivation for the
assumption that the radial pressure function (\ref{Eq14}) is
monotonically decreasing stems from the fact that, for
asymptotically flat hairy black-hole configurations, this pressure
function is known to be asymptotically decreasing \cite{Hodhair}. In
the present paper we shall prove that if this functional behavior is
assumed to hold true in the interval $[r_{\text{H}},r_{\gamma}]$,
then the null circular geodesics of the hairy black-hole spacetimes
are characterized by the physically interesting lower bound
(\ref{Eq2}).
\bibitem{Notehg} It is worth noting that our theorem would be valid for any matter theory in which the radial
pressure function $|r^3p(r)|$ decreases monotonically between the black-hole outer horizon and the innermost
null circular geodesic. In particular, the behavior of the pressure function in the spacetime region $r>r_{\gamma}$
outside the null circular geodesic is not important for our analysis.
\bibitem{Notenex} We consider non-extremal black holes for which ${\cal N}(r_{\text{H}})<0$ \cite{Hodhair}.
\end{thebibliography}
\end{document}